\documentclass[pre,showpacs,twocolumn]{revtex4}
\usepackage{graphicx}

\begin{document}

\title{Genetic attack on neural cryptography}
\date{1 December 2005}
\author{Andreas Ruttor}
\author{Wolfgang Kinzel}
\affiliation{Institut f\"ur Theoretische Physik, Universit\"at
  W\"urzburg, Am Hubland, 97074 W\"urzburg, Germany}
\author{Rivka Naeh}
\author{Ido Kanter}
\affiliation{Minerva Center and Department of Physics, Bar Ilan
  University, Ramat Gan 52900, Israel}
\begin{abstract}
  Different scaling properties for the complexity of bidirectional
  synchronization and unidirectional learning are essential for the
  security of neural cryptography. Incrementing the synaptic depth of
  the networks increases the synchronization time only polynomially,
  but the success of the geometric attack is reduced exponentially and
  it clearly fails in the limit of infinite synaptic depth. This
  method is improved by adding a genetic algorithm, which selects the
  fittest neural networks. The probability of a successful genetic
  attack is calculated for different model parameters using numerical
  simulations. The results show that scaling laws observed in the case
  of other attacks hold for the improved algorithm, too. The number of
  networks needed for an effective attack grows exponentially with
  increasing synaptic depth. In addition, finite-size effects caused
  by Hebbian and anti-Hebbian learning are analyzed. These learning
  rules converge to the random walk rule if the synaptic depth is
  small compared to the square root of the system size.
\end{abstract}
\pacs{84.35.+i, 87.18.Sn, 89.70.+c}
\maketitle

\section{Introduction}
\label{sec:intro}

Neural cryptography \cite{Kanter:2002:SEI, Kinzel:2003:DGI} is based
on the effect that two neural networks are able to synchronize by
mutual learning \cite{Metzler:2000:INN, Kinzel:2000:DIN}. In each step
of this online learning procedure they receive a common input pattern
and calculate their output. Then, both neural networks use those
outputs presented by their partner to adjust their own weights. So,
they act as teacher and student simultaneously. Finally, this process
leads to fully synchronized weight vectors.

Synchronization of neural networks is, in fact, a complex dynamical
process. The weights of the networks perform random walks, which are
driven by a competition of attractive and repulsive stochastic forces
\cite{Ruttor:2004:NCF}. Two neural networks can increase the
attractive effect of their moves by cooperating with each other. But,
a third network which is only trained by the other two clearly has a
disadvantage, because it cannot skip some repulsive steps. Therefore,
bidirectional synchronization is much faster than unidirectional
learning \cite{Kinzel:2003:DGI}.

This effect can be applied to solve a cryptographic problem: Two
partners $A$ and $B$ want to exchange a secret message. $A$ encrypts
the message to protect the content against an opponent $E$, who is
listening to the communication. But, $B$ needs $A$'s key in order to
decrypt the message. Therefore, the partners have to use a
cryptographic key-exchange protocol \cite{Stinson:1995:CTP} in order
to generate a common secret key. This can be achieved by synchronizing
two neural networks, one for $A$ and one for $B$, respectively. The
attacker $E$ trains a third neural network using inputs and outputs
transmitted by the partners as examples. But, on average, learning is
slower than synchronization. Thus, there is only a small probability
$P_E$ that $E$ is successful before $A$ and $B$ synchronize
\cite{Kinzel:2003:DGI}.

While other cryptographic algorithms use complicated calculations
based on number theory \cite{Stinson:1995:CTP}, the neural
key-exchange protocol only needs basic mathematical operations, namely
adding and subtracting integer numbers. These can be realized
efficiently in integrated circuits. Computer scientists are already
working on an hardware implementation of neural cryptography
\cite{Volkmer:2004:LCS, Volkmer:2005:KEI, Volkmer:2005:LKE,
  Volkmer:2005:TPM}.

Since the first proposal \cite{Kanter:2002:SEI} of the neural
key-exchange protocol, improved strategies for the attackers
\cite{Klimov:2003:ANC, Shacham:2003:CAN} and the partners
\cite{Mislovaty:2003:PCC, Ruttor:2004:NCF, Ruttor:2005:NCQ} have been
suggested and analyzed \cite{Kanter:2002:TNN, Kinzel:2002:INN,
  Kinzel:2003:DGI, Mislovaty:2002:SKE}. For the \emph{geometric
  attack} it has been found that the synaptic depth $L$ determines the
security of the system: the success probability $P_E$ decreases
exponentially with $L$, while the synchronization time
$t_\mathrm{sync}$ increases only proportionally to $L^2$
\cite{Mislovaty:2002:SKE, Ruttor:2004:SRW}. Therefore, any desired
level of security against this attack can be reached by increasing
$L$.

An improved version of this method is the \emph{majority attack}
\cite{Shacham:2003:CAN}. Here a group of $M$ neural networks estimates
the output of $B$'s hidden units. But, instead of updating the weights
individually, $E$'s tree parity machines cooperate and adjust the
weight vectors in the same way according to the majority vote. While
using this method increases $P_E$, the scaling laws hold except for
one special learning rule and random inputs \cite{Shacham:2003:CAN,
  Ruttor:2005:NCQ}. Therefore, neural cryptography is secure against
this attack in the limit $L \rightarrow \infty$, too.

In this paper we analyze a different method for the opponent E. The
\emph{genetic attack} \cite{Klimov:2003:ANC} is not based on optimal
learning like the majority attack \cite{Shacham:2003:CAN}, but employs
a genetic algorithm in order to select the most successful of $E$'s
neural networks. First, we repeat the definition of the neural
key-exchange protocol in Sec.~\ref{sec:neurocrypt}. We also explain
why $A$ and $B$ have a clear advantage over $E$. The algorithm of the
genetic attack is presented in Sec.~\ref{sec:attack}. Here, we show
that the scaling behavior observed for the geometric attack and the
majority attack also holds for the genetic attack. In
Sec.~\ref{sec:rules} we analyze the influence of the learning rules on
synchronization and learning. Finally, the known attacks on the neural
key-exchange protocol are compared regarding their efficiency. The
results presented in Sec.~\ref{sec:security} show that the genetic
attack is less efficient than the majority attack except for some
special cases.

\section{Neural cryptography}
\label{sec:neurocrypt}

\begin{figure}[tbp]
  \centering
  \includegraphics[width=8.6cm]{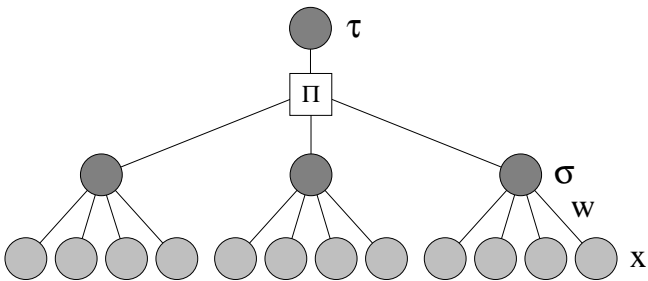}
  \caption{Tree parity machine with $K=3$ and $N=4$.}
  \label{fig:tpm}
\end{figure}

In this section we repeat the definition of the neural key-exchange
protocol \cite{Kanter:2002:SEI}. Each partner, $A$ and $B$, uses a
tree parity machine. The structure of this neural network is shown in
Fig.~\ref{fig:tpm}. A tree parity machine consists of $K$ hidden
units, which work like perceptrons. The possible input values are
binary,
\begin{equation}
  \label{eq:input}
  x_{ij} \in \{-1,+1\} \, ,
\end{equation}
and the weights are discrete numbers between $-L$ and $+L$,
\begin{equation}
  \label{eq:weight}
  w_{ij} \in \{-L,-L+1,\dots,L-1,L\} \, .
\end{equation}
Here the index $i=1,\dots,K$ denotes the $i$th hidden unit of the tree
parity machine and $j=1,\dots,N$ the elements in each vector. The
output of the first layer is defined as the sign of the scalar product
of inputs and weights,
\begin{equation}
  \label{eq:sigma}
  \sigma_i = \mathrm{sgn} \left( \mathbf{w}_i \cdot \mathbf{x}_i
  \right) \, .
\end{equation}
And, the total output of the tree parity machine is given by the
product (parity) of the hidden units,
\begin{equation}
  \label{eq:tau}
  \tau = \prod_{i=1}^K \sigma_i \, .
\end{equation}

At the beginning of the synchronization process $A$ and $B$ initialize
the weights of their neural networks randomly. This initial state is
kept secret. In each time step $t$, $K$ random input vectors
$\mathbf{x}_i$ are generated publicly and the partners calculate the
outputs $\tau^A$ and $\tau^B$ of their tree parity machines. After
communicating the output bits to each other they update the weights
according to one of the following learning rules:
\begin{enumerate}
  \renewcommand{\labelenumi}{(\roman{enumi})}
\item Hebbian learning
  \begin{equation}
    \label{eq:hp}
    \mathbf{w}_i^{+} = \mathbf{w}_i + \sigma_i \mathbf{x}_i
    \Theta(\sigma_i \tau) \Theta(\tau^A \tau^B) \, ,
  \end{equation}
\item Anti-Hebbian learning
  \begin{equation}
    \label{eq:hm}
    \mathbf{w}_i^{+} = \mathbf{w}_i - \sigma_i \mathbf{x}_i
    \Theta(\sigma_i \tau) \Theta(\tau^A \tau^B) \, ,
  \end{equation}
\item Random walk
  \begin{equation}
    \label{eq:rw}
    \mathbf{w}_i^{+} = \mathbf{w}_i + \mathbf{x}_i \Theta(\sigma_i
    \tau) \Theta(\tau^A \tau^B) \, .
  \end{equation}
\end{enumerate}
If any component of the weight vectors moves out of the range
$-L,\dots,+L$, it is replaced by the nearest boundary value, either
$-L$ or $+L$.

After some time $t_\mathrm{sync}$ the partners have synchronized their
tree parity machines, $\mathbf{w}_i^A(t_\mathrm{sync}) =
\mathbf{w}_i^B(t_\mathrm{sync})$, and the process is stopped.
Afterwards, $A$ and $B$ can use the weight vectors as a common secret
key in order to encrypt and decrypt secret messages.

We describe the process of synchronization by standard order
parameters, which are also used for the analysis of online learning
\cite{Engel:2001:SML}. These order parameters are
\begin{eqnarray}
  \label{eq:qrdef}
  Q_i^m     &=& \frac{1}{N} \mathbf{w}_i^m \cdot \mathbf{w}_i^m \, , \\
  R_i^{m,n} &=& \frac{1}{N} \mathbf{w}_i^m \cdot \mathbf{w}_i^n \, ,
\end{eqnarray}
where the indices $m,n \in \{A,B,E\}$ denote $A$'s, $B$'s or $E$'s
tree parity machine, respectively. The level of synchronization
between two corresponding hidden units is defined by the (normalized)
overlap,
\begin{equation}
  \label{eq:rho}
  \rho_i^{m,n} = \frac{\mathbf{w}_i^m \cdot
    \mathbf{w}_i^n}{\sqrt{\mathbf{w}_i^m \cdot \mathbf{w}_i^m}
    \sqrt{\mathbf{w}_i^n \cdot \mathbf{w}_i^n}} =
  \frac{R_i^{m,n}}{\sqrt{Q_i^m Q_i^n}} \, .
\end{equation}
Uncorrelated weight vectors have $\rho=0$, while the maximum value
$\rho=1$ is reached for full synchronization.

The overlap between two corresponding hidden units increases if the
weights of both neural networks are updated in the same way.
Coordinated moves, which occur for identical $\sigma_i$, have an
attractive effect.

Changing the weights in only one hidden unit decreases the overlap on
average. These repulsive steps can only occur if the two output values
$\sigma_i$ are different. The probability for this event is given by
the well-known generalization error of the perceptron,
\cite{Engel:2001:SML}
\begin{equation}
  \label{eq:epsilon}
  \epsilon_i = \frac{1}{\pi} \arccos \rho_i \, ,
\end{equation}
which itself is a function of the overlap $\rho_i$ between the hidden
units. For an attacker who simply trains a third tree parity machine
using the examples generated by $A$ and $B$, repulsive steps occur
with probability $P_r^E=\epsilon_i$, because $E$ cannot influence the
process of synchronization.

In contrast, $A$ and $B$ communicate with each other and are able to
interact. If they disagree on the total output, there is at least one
hidden unit with $\sigma_i^A \neq \sigma_i^B$. As an update would have
a repulsive effect, the partners just do not change the weights. In
doing so, $A$ and $B$ reduce the probability of repulsive steps in
their hidden units. For $K=3$ and identical generalization error,
$\epsilon_i=\epsilon$, we find \cite{Ruttor:2004:NCF}
\begin{equation}
  \label{eq:rep}
  P_r^B = \frac{2 (1 - \epsilon) \epsilon^2}{(1 - \epsilon)^3 + 3 (1 -
    \epsilon) \epsilon^2} \leq \epsilon = P_r^E \, .
\end{equation}
Therefore, the partners have a clear advantage over an attacker using
only simple learning.

But, $E$ can use a more advanced method called \emph{geometric
  attack}. As before, she trains a third tree parity machine, which
has the same structure as $A$'s and $B$'s. In each step $\tau^E$ is
calculated and compared to $\tau^B$. As long as these output values
are identical, $E$ can apply the learning rule in the same manner as
$B$. But, if $\tau^E \neq \tau^B$, the attacker has to correct this
deviation before updating the weights.

For this purpose $E$ uses the local field
\begin{equation}
  \label{eq:field}
  h_i = \frac{1}{\sqrt{N}} \mathbf{w}_i \cdot \mathbf{x}_i
\end{equation}
of her hidden units as additional information. Then, the probability
of $\sigma_i^B \neq \sigma_i^E$ is given by the prediction error of
the perceptron \cite{Ein-Dor:1999:CPN}
\begin{equation}
  \label{eq:epsilon2}
  \epsilon_i(\rho_i,h_i) = \frac{1}{2} \left[ 1 - \mathrm{erf} \left(
      \frac{\rho_i}{\sqrt{2(1-\rho_i^2)}} \frac{|h_i|}{\sqrt{Q_i}}
    \right) \right] \, .
\end{equation}
If the local field $h_i$ is zero, the neural network has no
information about the input vector $\mathbf{x}_i$, because it is
perpendicular to the weight vector $\mathbf{w}_i$. In this case the
prediction error reaches its global maximum of
$\epsilon_i(\rho_i,0)=1/2$.

\begin{figure}[tbp]
  \centering
  \includegraphics[width=8.6cm]{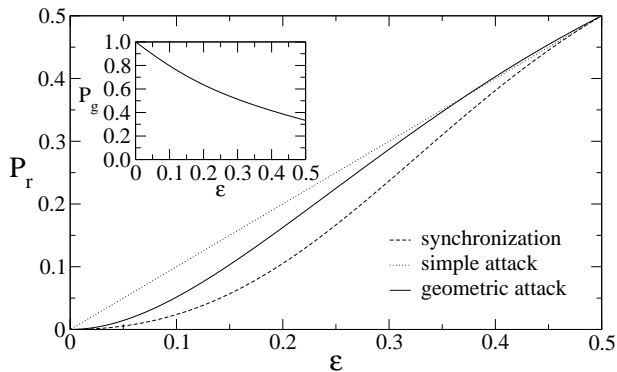}
  \caption{Probability $P_r$ of repulsive steps as a function of the
    generalization error $\epsilon$. The inset shows the probability
    $P_g$ for a successful geometric correction.}
  \label{fig:prcmp}
\end{figure}

The prediction error $\epsilon_i(\rho_i,h_i)$ is a strictly monotonic
decreasing function of $|h_i|$. Therefore, the attacker searches the
hidden unit with the lowest value of the absolute local field
$|h_i^E|$ and flips the sign of $\sigma_i^E$. This results in $\tau^E
= \tau^B$ and the learning rule can be applied. But, the geometric
attack does not always find the correct hidden unit which caused the
deviation of the total output bits. If $\sigma_i^E \neq \sigma_i^B$ in
the $i$th hidden unit and $\sigma_j^E = \sigma_j^B$ in all other
hidden units, $E$ flips the sign of $\sigma_i^E$ with probability
\begin{eqnarray}
  \label{eq:pg}
  P_g &=& \int_{0}^{\infty} \prod_{j \neq i} \left(
    \int_{h_i}^{\infty} \sqrt{\frac{2 \pi}{Q_j}} \frac{1 -
      \epsilon_j(\rho_j,h_j)}{\pi - \arccos \rho_j} e^{-\frac{h_j^2}{2
        Q_j}} \mathrm{d}h_j \right) \nonumber \\
  &\times& \sqrt{\frac{2 \pi}{Q_i}}
  \frac{\epsilon_i(\rho_i,h_i)}{\arccos \rho_i} e^{-\frac{h_i^2}{2
      Q_i}} \mathrm{d}h_i \, .
\end{eqnarray}
Thus, the geometric attacker avoids some repulsive steps, although
they still occur more frequently than in the partners' tree parity
machines.

In the case of identical generalization error $\epsilon_i=\epsilon$
and $K=3$, we find that the probability of repulsive steps,
\begin{equation}
  \label{eq:pgr}
  P_r^E = 2 (1 - P_g) (1 - \epsilon)^2 \epsilon + 2 (1 - \epsilon)
  \epsilon^2 + \frac{2}{3} \epsilon^3 \, ,
\end{equation}
is higher than $P_r^B$, but lower than $P_r^E = \epsilon$ for simple
learning. This result is clearly visible in Fig.~\ref{fig:prcmp}.
That is why learning by listening is slower than mutual learning, even
for advanced algorithms. This effect makes neural cryptography
feasible and prevents successful attacks in the limit $L \rightarrow
\infty$.

Recently, it has been discovered that the security of the neural
key-exchange protocol can be improved by using queries instead of
random inputs \cite{Kinzel:1990:ING, Ruttor:2005:NCQ}. The partners
ask questions to each other which depend on their own weight vectors
$\mathbf{w}_i$ and an additional public parameter $H$. In odd (even)
steps $A$ ($B$) generates $K$ input vectors $\mathbf{x}_i$ with $h_i^A
\approx \pm H$ ($h_i^B \approx \pm H$). So, the absolute value of the
local field $h_i$ is given by $H$, while its sign $\sigma_i$ is chosen
randomly.

\begin{figure}[tbp]
  \centering
  \includegraphics[width=8.6cm]{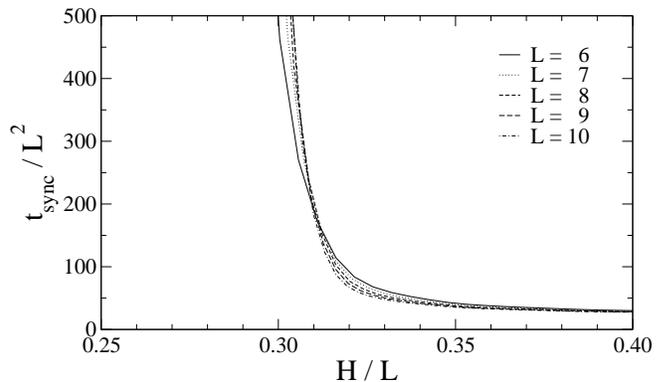}
  \caption{Synchronization time $t_\mathrm{sync}$ as a function of $H$
    for $K=3$, $N=1000$, random walk learning rule, and different
    values of $L$, averaged over $10\,000$ simulations.}
  \label{fig:sync}
\end{figure}

Queries change the relation between the overlap and the frequency
$P_r$ of repulsive steps. The probability of different outputs
$\sigma_i$ in corresponding hidden units is now given by
Eq.~(\ref{eq:epsilon2}) instead of Eq.~(\ref{eq:epsilon}), because the
absolute local field in $A$'s or $B$'s hidden units is known.
Consequently, the partners can optimize complexity and security of the
neural key-exchange protocol by adjusting $H$ and $L$ suitably
\cite{Ruttor:2005:NCQ}.

As shown in Fig.~\ref{fig:sync}, a minimum value of $H$ is needed in
order to achieve synchronization in a reasonable number of steps. If
$H > \alpha_c L$, $t_\mathrm{sync}$ increases proportional to $L^2 \ln
N$, but for $H < \alpha_c L$ it diverges
\cite{Ruttor:2005:NCQ,Ruttor:2004:SRW}. In the case of the random walk
learning rule we estimate $\alpha_c \approx 0.31$ by using the
extrapolation method described in \cite{Ruttor:2005:NCQ}.

\section{Genetic Attack}
\label{sec:attack}

\begin{figure}[tbp]
  \centering
  \includegraphics[width=8.6cm]{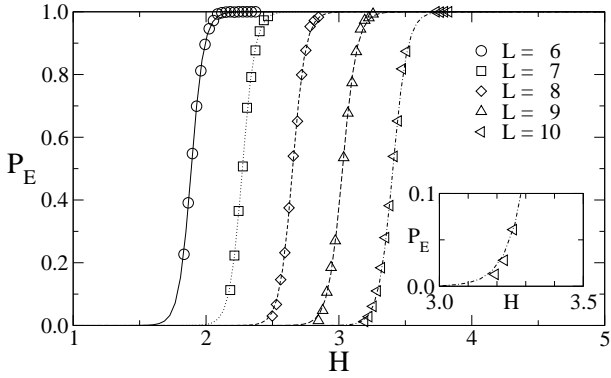}
  \caption{Success probability $P_E$ of the genetic attack for $K=3$,
    $N=1000$, random walk learning rule, and $M=4096$. Symbols
    represent results obtained from $1000$ simulations, and the lines
    show a fit with Eq.~(\ref{eq:fit}).}
  \label{fig:success}
\end{figure}

For the genetic attack \cite{Klimov:2003:ANC} the opponent starts with
only one tree parity machine, but she can use up to $M$ neural
networks. As before $E$ calculates the output of her networks in each
step. Afterwards the following genetic algorithm is applied:
\begin{enumerate}
  \renewcommand{\labelenumi}{(\roman{enumi})}
\item If $\tau^A = \tau^B$ and $E$ has at most $M / 2^{K-1}$ tree
  parity machines, she determines all $2^{K-1}$ internal
  representations $(\sigma_1^E,\dots,\sigma_K^E)$ which reproduce the
  output $\tau^A$. Then, these are used to update the weights in $E$'s
  neural networks according to the learning rule, so that $2^{K-1}$
  variants of each tree parity machine are generated.
\item But, if $E$ already has more than $M / 2^{K-1}$ neural networks,
  the \emph{mutation step} described above is not possible. Instead of
  that the attacker discards all tree parity machines which predicted
  less than $U$ outputs $\tau^A$ in the last $V$ learning steps, with
  $\tau^A = \tau^B$, successfully. In our simulations we use a limit
  of $U=10$ and a history of $V=20$ as default values. Additionally,
  at least 20 neural networks are kept in such a \emph{selection
    step}.
\item In the case of $\tau^A \neq \tau^B$ the attacker's networks
  remain unchanged, because $A$ and $B$ do not update the weights in
  their tree parity machines.
\end{enumerate}
The attack is considered successful if at least one of $E$'s neural
networks has synchronized $98\%$ of the weights before the end of the
key exchange. We use this relaxed criterion in order to decrease the
fluctuations of $P_E$ \cite{Mislovaty:2002:SKE}.

The success probability of the genetic attack strongly depends on the
value of the parameter $H$. This effect is clearly visible in
Fig.~\ref{fig:success}. In order to determine $P_E$ as a function of
$H$, a Fermi-Dirac distribution
\begin{equation}
  \label{eq:fit}
  P_E = \frac{1}{1 + \exp[-\beta (H - \mu)]}
\end{equation}
with two parameters $\beta$ and $\mu$ can be used as a fitting model.
This equation is also valid for the geometric attack and the majority
attack \cite{Ruttor:2005:NCQ}.

\begin{figure}[tbp]
  \centering
  \includegraphics[width=8.6cm]{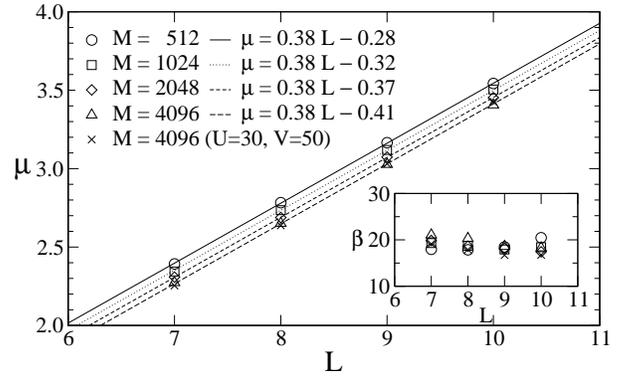}
  \caption{Parameters $\mu$ and $\beta$ as a function of the synaptic
    depth $L$. Symbols denote results of fitting simulation data for
    different $M$ with Eq.~(\ref{eq:fit}) and the lines were
    calculated using the model given in Eq.~(\ref{eq:gamma}).}
  \label{fig:attack1}
\end{figure}

\begin{figure}[tbp]
  \centering
  \includegraphics[width=8.6cm]{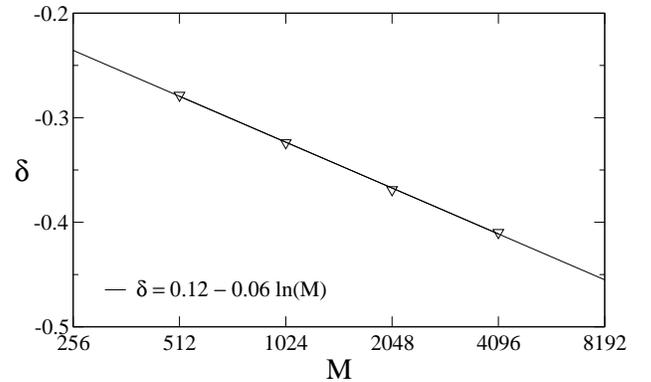}
  \caption{Offset $\delta$ as a function of the number of attackers
    $M$, for $K=3$, $N=1000$, and the random walk learning rule.
    Symbols and the line were obtained by a fit with
    Eq.~(\ref{eq:gamma}).}
  \label{fig:attack2}
\end{figure}

Figure~\ref{fig:attack1} shows the results of the fits using
Eq.~(\ref{eq:fit}). While $\beta$ is nearly independent of $L$ and
$M$, $\mu$ increases linearly with the synaptic depth,
\begin{equation}
  \label{eq:mu}
  \mu = \alpha_s L + \delta \, .
\end{equation}
Obviously, the attacker can change the offset $\delta$, but not
$\alpha_s$, by using more resources. As shown in
Fig.~\ref{fig:attack2} $E$ needs to double $M$ in order to decrease
$\delta$ by a fixed amount $\gamma \ln 2$. Thus, $\mu$ is a linear
function of both $L$ and $\ln M$,
\begin{equation}
  \label{eq:gamma}
  \mu = \alpha_s L - \gamma \ln M + \mu_0 \, .
\end{equation}
Substituting Eq.~(\ref{eq:gamma}) into Eq.~(\ref{eq:fit}) leads to
\begin{equation}
  \label{eq:pe}
  P_E = \frac{1}{1 + \exp[\beta (\mu_0 - \gamma \ln M)] \exp[\beta
    (\alpha_s - \alpha) L]}
\end{equation}
for the success probability of the genetic attack as a function of
$\alpha = H / L$, the synaptic depth $L$, and the maximal number of
attackers $M$.

From these results we can deduce the scaling of $P_E$ with regard to
$L$ and $M$. For large values of the synaptic depth the asymptotic
behavior is given by
\begin{equation}
  \label{eq:scale1}
  P_E \sim e^{-\beta (\mu_0 - \gamma \ln M)} e^{-\beta
    (\alpha_s - \alpha) L}
\end{equation}
as long as $\alpha < \alpha_s$.

\begin{figure}[tbp]
  \centering
  \includegraphics[width=8.6cm]{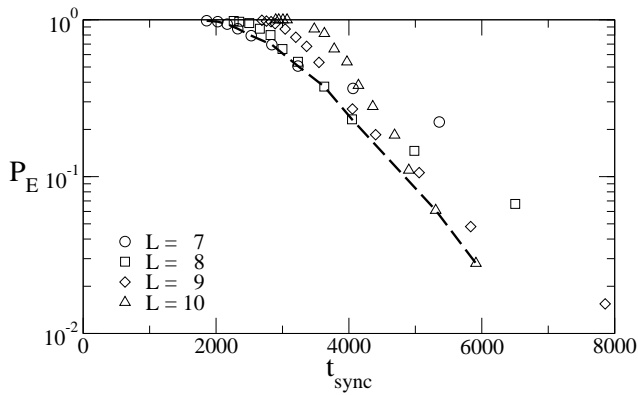}
  \caption{Success probability of the \emph{genetic attack} as a
    function of the synchronization time for $K=3$, $N=1000$, random
    walk learning rule, $M=4096$ and different values of $L$. The
    dashed line shows the envelope of this set of curves.}
  \label{fig:result}
\end{figure}

This equation shows that that the partners have a great advantage over
an attacker. If $A$ and $B$ increase $L$, the success probability
drops exponentially,
\begin{equation}
  \label{eq:scale2}
  P_E \propto e^{-\beta (\alpha_s - \alpha) L} \, ,
\end{equation}
while the complexity of the synchronization rises only polynomially.
This is clearly visible if one looks at the function $P_E(\langle
t_\mathrm{sync} \rangle)$, which is shown in Fig.~\ref{fig:result}.
Due to the offset $\delta$ in Eq.~(\ref{eq:mu}) the attacker is
successful for small values of $L$. But, for larger synaptic depth
optimal security is reached for values of $H$ and $L$, which lie on
the envelope of $P_E(\langle t_\mathrm{sync} \rangle)$. This curve is
approximately given by $H = \alpha_c L$, as this condition maximizes
$\alpha_s - \alpha$ while synchronization is still possible
\cite{Ruttor:2005:NCQ}.

In contrast, the attacker has to increase the number of her tree
parity machines exponentially,
\begin{equation}
  \label{eq:scale3}
  M \propto e^{[(\alpha_s - \alpha) / \gamma] L} \, ,
\end{equation}
in order to compensate a change of $L$ and maintain a constant success
probability $P_E$. But, this is usually not possible due to limited
computer power.

\begin{figure}[tbp]
  \centering
  \includegraphics[width=8.6cm]{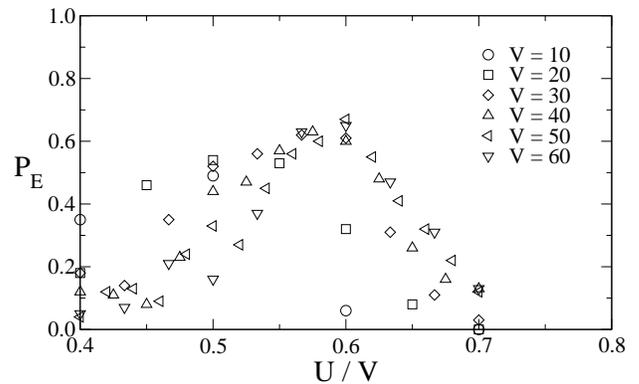}
  \caption{Success probability of the genetic attack for $K=3$, $L=7$,
    $N=1000$, random walk learning rule, $M=4096$, and $H=2.28$. These
    results were obtained by averaging over $100$ simulations.}
  \label{fig:gwh2}
\end{figure}

Alternatively, the attacker could try to optimize the other two
parameters of the genetic attack. As shown in Fig.~\ref{fig:gwh2}, $E$
obtains the best result if she uses $U=30$, $V=50$ instead of $U=10$,
$V=20$. Figure~\ref{fig:attack1} shows that this modification leads to
a lower value of $\beta$, but does not influence $\mu(L)$. Therefore,
$E$ gains little, as the scaling relation (\ref{eq:scale3}) is not
affected. That is why $A$ and $B$ can easily reach an arbitrary level
of security.

\section{Learning rules}
\label{sec:rules}

Beside the random walk learning rule (\ref{eq:rw}) used so far, there
are two other suitable algorithms for updating the weights: the
Hebbian learning rule (\ref{eq:hp}) and the anti-Hebbian learning rule
(\ref{eq:hm}). The only difference between these three rules is
whether and how the output $\sigma_i$ of the hidden unit is included
in the update step. But, this causes some effects which we discuss in
this section.

In the case of the Hebbian rule $A$'s and $B$'s tree parity learn
their own output. Therefore, the direction in which the weight
$w_{ij}$ moves is determined by the product $\sigma_i x_{ij}$. But, as
the output of a hidden units is a function of all input values, there
are correlations between $x_{ij}$ and $\sigma_i$. That is why the
probability distribution of $\sigma_i x_{ij}$ is not uniformly
distributed in the case of random inputs, but depends on the
corresponding weight $w_{ij}$,
\begin{equation}
  \label{eq:corr}
  P(\sigma_i x_{ij} = 1) = \frac{1}{2} \left[ 1 + \mathrm{erf} \left(
      \frac{w_{ij}}{\sqrt{N Q_i - w_{ij}^2}} \right) \right] \, .
\end{equation}
According to this equation, $\sigma_i x_{ij} = \mathrm{sgn}(w_{ij})$
occurs more often than $\sigma_i x_{ij} = -\mathrm{sgn}(w_{ij})$.
Thus, the Hebbian learning rule (\ref{eq:hp}) pushes the weights
towards the boundaries at $-L$ and $+L$.

In order to quantify this effect we calculate the stationary
probability distribution of the weights. Using Eq.~(\ref{eq:corr}) for
the transition probabilities leads to
\begin{equation}
  \label{eq:pw}
  P(w_{ij} = w) = p_0 \prod_{m=1}^{|w|} \frac{1 + \mathrm{erf} \left(
      \frac{m - 1}{\sqrt{N Q_i - (m - 1)^2}} \right)}{1 - \mathrm{erf}
    \left( \frac{m}{\sqrt{N Q_i - m^2}} \right)} \, ,
\end{equation}
whereas the normalization constant $p_0$ is given by
\begin{equation}
  \label{eq:norm}
  p_0 = \left( \sum_{w = -L}^{L} \prod_{m=1}^{|w|} \frac{1 +
      \mathrm{erf} \left( \frac{m - 1}{\sqrt{N Q_i - (m - 1)^2}}
      \right)}{1 - \mathrm{erf} \left( \frac{m}{\sqrt{N Q_i - m^2}}
      \right)} \right)^{-1} \, .
\end{equation}
In the limit $N \rightarrow \infty$ the argument of the error function
vanishes and the weights are uniformly distributed. In this case the
synchronization process does not change the initial length
\begin{equation}
  \label{eq:length}
  \sqrt{Q_i(t = 0)} = \sqrt{\frac{L(L+1)}{3}}
\end{equation}
of the weight vector.

\begin{figure}[tbp]
  \centering
  \includegraphics[width=8.6cm]{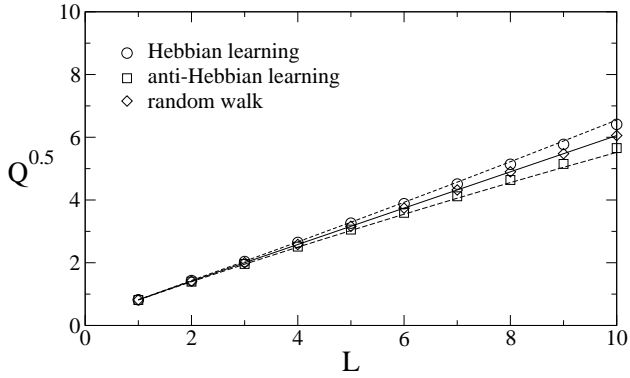}
  \caption{Length of the weight vectors in the steady state
    for $K=3$ and $N=1000$. Symbols denote results averaged over
    $1000$ simulations and lines show the first-order approximation
    given in Eq.~(\ref{eq:q2}) and Eq.~(\ref{eq:q4}).}
  \label{fig:length}
\end{figure}

But, for finite $N$ the probability distribution (\ref{eq:pw}) itself
depends on the order parameter $Q_i$. Therefore, the expectation value
of $Q_i$ is the solution of the following equation:
\begin{equation}
  \label{eq:condition}
  Q_i = \sum_{w = -L}^{L} w^2 P(w_{ij} = w) \, .
\end{equation}
By expanding Eq.~(\ref{eq:condition}) in terms of $N^{-1/2}$ we obtain
\begin{eqnarray}
  \label{eq:q2}
  Q_i &=& \frac{L (L + 1)}{3} \nonumber \\
  &+& \frac{8 L^4 + 16 L^3 -10 L^2 - 18 L + 9}{15 \sqrt{3 \pi L (L +
      1)}} \frac{1}{\sqrt{N}} \nonumber \\
  &+& \mathrm{O} \left( \frac{L^4}{N} \right)
\end{eqnarray}
as a first-order approximation of $Q_i$ for large system sizes. In the
case of $1 \ll L \ll \sqrt{N}$ the asymptotic behavior of this order
parameter is given by
\begin{equation}
  \label{eq:q3}
  Q_i \sim \frac{L (L + 1)}{3} \left( 1 + \frac{8}{5 \sqrt{3 \pi}}
    \frac{L}{\sqrt{N}} \right) \, .
\end{equation}
Obviously, the application of the Hebbian learning rule increases the
length of the weight vectors $\mathbf{w}_i$ until a steady state is
reached. Additionally, the changed probability distribution of the
weights affects the synchronization process and the success of
attacks. That is why one encounters finite-size effects if $L /
\sqrt{N}$ is large \cite{Mislovaty:2002:SKE}.

In the case of the anti-Hebbian rule $A$'s and $B$'s tree parity
machines learn the opposite of their own outputs. Therefore, the
weights are pulled away from the boundaries, so that
\begin{eqnarray}
  \label{eq:q4}
  Q_i &=& \frac{L (L + 1)}{3} \nonumber \\
  &-& \frac{8 L^4 + 16 L^3 - 10 L^2 - 18 L + 9}{15 \sqrt{3 \pi L (L +
      1)}} \frac{1}{\sqrt{N}} \nonumber \\
  &+& \mathrm{O} \left( \frac{L^4}{N} \right) \\
  \label{eq:q5}
  &\sim& \frac{L (L + 1)}{3} \left( 1 - \frac{8}{5 \sqrt{3 \pi}}
    \frac{L}{\sqrt{N}} \right)
\end{eqnarray}
for $1 \ll L \ll \sqrt{N}$. Here, the length of the weight vectors
$\mathbf{w}_i$ is decreased.

In contrast, the random walk learning rule always uses a fixed set
output. Here, the weights stay uniformly distributed, because only the
random input values $x_{ij}$ determine the direction of the movements.
In this case the length of the weight vectors is given by
Eq.~(\ref{eq:length}).

\begin{figure}[tbp]
  \centering
  \includegraphics[width=8.6cm]{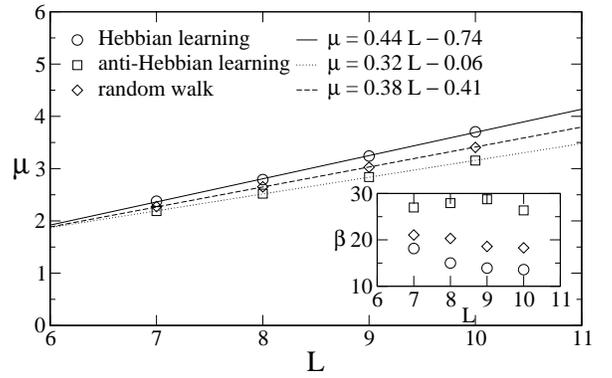}
  \caption{Parameter $\mu$ and $\beta$ as a function of $L$ for the
    genetic attack with $K=3$, $N=1000$, and $M=4096$. The symbols
    represent results from $1000$ simulations and the lines show a fit
    using the model given in Eq.~(\ref{eq:gamma}).}
  \label{fig:attack3}
\end{figure}

Figure~\ref{fig:length} shows that the theoretical predictions are in
good quantitative agreement with simulation results as long as $L^2$
is small compared to the system size $N$. The deviations for large $L$
are caused by higher-order terms which are ignored in
Eq.~(\ref{eq:q2}) and Eq.~(\ref{eq:q4}).

\begin{figure}[tbp]
  \centering
  \includegraphics[width=8.6cm]{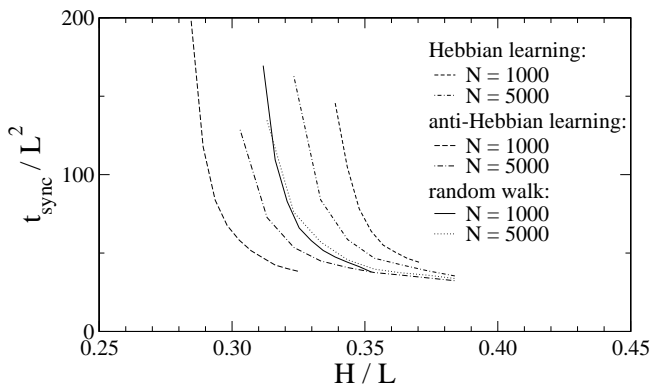}
  \caption{Synchronization time $t_\mathrm{sync}$ as a function of $H$
    for $K=3$, $L=7$, averaged over $100$ simulations.}
  \label{fig:sync2}
\end{figure}

The choice of the learning rule affects synchronization with random
inputs as well as with queries. As the prediction error
(\ref{eq:epsilon2}) is a function of $h_i / \sqrt{Q_i}$, this ratio
instead of just the local field determines the behavior of the system.
That is why there are different values of $\alpha_c$ and $\alpha_s$
for each learning rule, which is shown in Fig.~\ref{fig:attack3} and
Fig.~\ref{fig:sync2}.

In the limit $N \rightarrow \infty$, however, a system using Hebbian
or anti-Hebbian learning exhibits the same dynamics as observed in the
case of the random walk rule for all system sizes. This is clearly
visible in Fig.~\ref{fig:sync2}. Consequently, one can determine the
properties of neural cryptography in the limit $N \rightarrow \infty$
without actually analyzing very large systems. It is sufficient to use
the random walk learning rule and moderate values of $N$ in
simulations.

\section{Security}
\label{sec:security}

In order to assess the security of the neural key-exchange protocol
one has to consider all known attacks. Therefore, we compare the
efficiency of several methods here.

\begin{figure}[tbp]
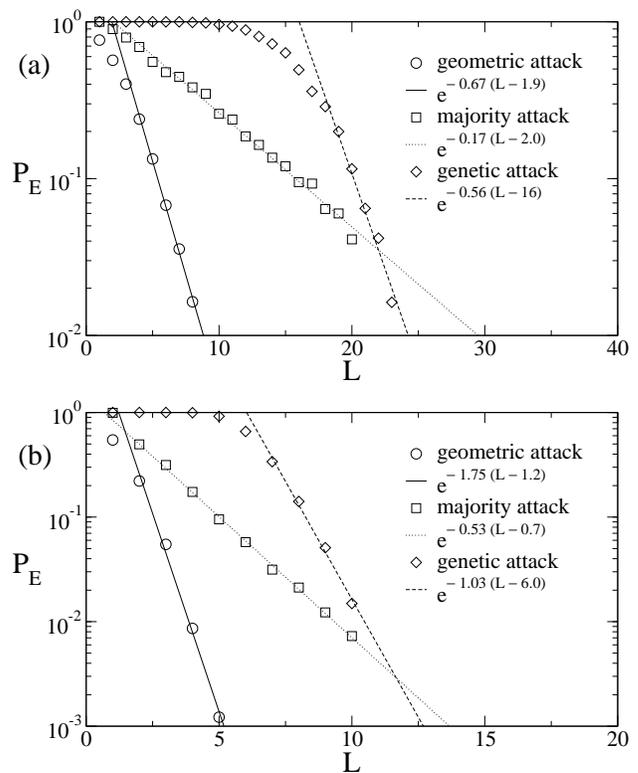

  \centering
  \includegraphics[width=8.6cm]{random.eps}
  \includegraphics[width=8.6cm]{queries.eps}
  \caption{Success probability $P_E$ of the geometric attack with
    $M=1$, the majority attack with $M=100$, and the genetic attack
    with $M=4096$, for $K=3$, $N=1000$, and the random walk learning
    rule. Symbols represent results averaged over $1000$ simulations,
    in part (a) for random inputs and in part~(b) for queries with
    $H=0.32 L$. The lines were obtained by fitting with
    Eq.~(\ref{eq:pe2}).}
  \label{fig:result2}
\end{figure}

Figure~\ref{fig:result2} shows that the success probability $P_E$
drops exponentially with increasing synaptic depth $L$,
\begin{equation}
  \label{eq:pe2}
  P_E \sim e^{- y (L - L_0)} \, ,
\end{equation}
as long as $L > L_0$. While this scaling behavior is the same for all
attacks, the constants $y$ and $L_0$ are different for each method.

The geometric attack is the simplest method considered here. $E$ only
needs one tree parity machine, but the success probability $P_E$ is
lower than for the advanced methods. As the exponent $y$ is large, the
two partners can easily secure the neural key-exchange protocol by
increasing the synaptic depth \cite{Mislovaty:2002:SKE}.

In the case of the majority attack $P_E$ is higher, because the
cooperation between $E$'s tree parity machines reduces the coefficient
$y$. $A$ and $B$ have to compensate this by further stepping up $L$.
In contrast, the genetic attack increases $L_0$, while $y$ does not
change significantly compared to the geometric attack. Therefore, the
genetic algorithm is better only if $L$ is not too large. Otherwise
$E$ gains most by using the majority attack.

As shown in Fig.~\ref{fig:result2} the partners can improve the
security of the key-exchange protocol against all three attacks by
using queries. However, the majority attack remains the most efficient
of $E$'s methods.

We note that these results are based on numerical extrapolations of
the success probability $P_E$. While analytical evidence for the
complexity of a successful attack would be desirable, it is not
available yet in the case of the nondeterministic methods with $P_E<1$
discussed above. But there are only two successful deterministic
algorithms for $E$ known at present: a brute-force attack or a genetic
attack with $M=2^{(K-1) t_\mathrm{sync}}$ networks. The complexity of
these attacks clearly grows exponentially with increasing $L$.
Therefore, breaking the security of neural cryptography belongs to the
complexity class $NP$ (nondeterministic polynomial time), but we
cannot prove that it is not in $P$ (polynomial time). This situation
is similar to that of other cryptographic protocols, e.g., the
Diffie-Hellman key exchange \cite{Stinson:1995:CTP}.

\section{Conclusions}
\label{sec:cc}

The security of cryptographic algorithms is usually based on different
scaling laws regarding the computational complexity for users and
attackers. By changing some parameter one can increase the cost of a
successful attack exponentially, while the effort for the users
increments only polynomially. For conventional cryptographic systems
this parameter is the length of the key. In the case of neural
cryptography it is the synaptic depth $L$ of the neural networks.

As the neural key-exchange protocol uses tree parity machines, an
attacker faces the challenge to guess the internal representation of
these networks correctly. Learning alone is not sufficient to solve
this problem. Otherwise the scaling laws hold and the partners can
achieve any desired level of security by increasing $L$.

We have analyzed an attack, which combines learning with a genetic
algorithm. We have found that this method is very successful as long
as $L$ is small. But, attackers have to increase the number of their
neural networks exponentially in order to compensate higher values of
$L$. That is why neural cryptography is secure against the genetic
attack as well.

This method achieves the best success probability of all known methods
only if the synaptic depth $L$ is not too large. For higher values of
$L$ the attacker gains more by using the majority attack. But, both
methods are unable to break the security of the neural key-exchange
protocol in the limit $L \rightarrow \infty$.

Additionally, we have studied the influence of different learning
rules on the neural key-exchange protocol. Hebbian and anti-Hebbian
learning change the order parameter $Q$, which is related to the
length of the weight vectors. If the system size $N$ is small compared
to $L^2$, this causes finite-size effects. But, in the limit $L /
\sqrt{N} \rightarrow 0$ the behavior of all learning rules converges
to that of the random walk rule.

Based on our results, we conclude that the neural key-exchange
protocol is secure against all attacks known up to now. But---similar
to other cryptographic algorithms---there is always a possibility that
a clever method may be found which destroys the security of neural
cryptography completely.

\bibliography{paper}

\end{document}